\documentclass[10pt]{iopart}
\usepackage{graphicx}

\begin{document}

\title{Analysis of the finite-size effect of the long-range Ising model under Glauber dynamics}

\author{Hisato Komatsu } 
\address{ Interdisciplinary Graduate School of Engineering Sciences, Kyushu University, Kasuga, Fukuoka 816-8580, Japan}


\begin{abstract}
We considered a long-range Ising model under Glauber dynamics and calculated the difference from the mean-field approximation in a finite-size system using perturbation theory. To deal with the BBGKY hierarchy, we assumed that certain types of extensive properties have a Gaussian distribution, which turned out to be equivalent to the Kirkwood superposition approximation within the range of first-order perturbation. After several calculations, ordinary differential equations that describe the time development of a two-body correlation were derived.
This discussion is the generalization of our previous study which developed a similar consideration on the infinite-range Ising model. The results of the calculation fit those of the numerical simulations for the case in which the decay of the interaction was sufficiently slow; however, they exhibited different behaviors when the decay became rapid.
\end{abstract}

\maketitle

\section{Introduction \label{Introduction} }

Long-range interactions play significant roles in numerous fields, such as plasma and astronomical physics \cite{CDFR14,CDR09,LPRTB14}. To deal with these interactions, a pair of variables that are spatially distant from each other should be considered. Therefore, studies on these systems are often difficult. However, in certain cases, theoretical approaches which cannot be used for usual short-range interaction systems can be adopted.
 For example, the mean-field approximation tends to be more accurate for long-range interaction systems than short-range systems. This ``exactness of mean-field theory'' has been extensively studied with both a system in an equilibrium \cite{TA00,CMT00,CGM00,Mori11,Mori12} and nonequilibrium state \cite{DOPT94,DOPT96_1,DOPT96_2,BDNRS11}. Most of these previous studies focused on the behavior of the thermodynamic limit; however, investigations into finite-sized systems have recently begun \cite{Kastner11,Mori19}.
 
The finite-size effect plays an important role in various statistical physics problems. In particular, in the case of long-range interaction systems, numerical simulations require longer computational times than short-range systems. Furthermore, certain types of long-range interaction systems exhibit divergence of the relaxation time in the thermodynamic limit \cite{Kastner11,BBDRY06,Komatsu20}. Hence, studies on large systems are considerably more difficult, and the finite-size effect must be considered to understand the results of the simulations.

As one of the simplest examples, an infinite-range Ising model under Glauber dynamics has previously been studied \cite{PHB89,AFC10,MMR10,GMM11}. Most of these studies considered the probability density function of magnetization, $P_{\mathrm{whole} } (m )$, and derived the Fokker--Planck equation to describe the time development of this function. 
However, their discussions required information on the number of microscopic states under fixed-order parameters; hence, they cannot be easily generalized to other systems.

In our previous study, we considered another method of discussing the finite-size effect of the infinite-range Ising model, treating the difference from the mean-field approximation as perturbation \cite{Komatsu22}. 
In this study, we generalized this method to the long-range Ising model, which has an interaction exhibiting power-law decay. The fundamental methodology is similar to that of our previous study. We first considered the probability distribution of two spin variables $P_{2,i,j} ( \sigma _i , \sigma _j ; t )$, regarded the difference from the mean-field approximation as perturbation, and calculated the time-development of this quantity. During the calculation, we encountered a problem known as BBGKY hierarchy \cite{CDFR14}. To deal with this problem, the probability distributions of extensive properties with specific forms were assumed to be Gaussian, although this assumption is difficult to verify. As explained later, this assumption turned out to be equivalent to the Kirkwood superposition approximation within the range of first-order perturbation \cite{Kirkwood42,CL64,AJM97}. 
As a result of the calculation, we obtained the ordinary differential equation describing the time development of the magnetization and the spatial correlation. Note that in the case of the Ising model, calculation of these macroscopic quantities is equivalent to that of the probability distribution $P_2$ itself. 

To confirm the validity of perturbation theory and the assumption, the results of the derived differential equations were compared with those of numerical simulations. According to this comparison, our discussion became accurate when the decay of the interaction was sufficiently slow and the system size $N$ was sufficiently large.

 The remainder of this paper is organized as follows: First, we explain the model in Section \ref{Model}. Then, Section \ref{Calculation} contains the calculation of the time development of the system using perturbation theory, and a comparison of the results with those of the numerical simulation are presented in Section \ref{Simulation}. Finally, the study is summarized in Section \ref{Summary}. We also explain the reason why the assumption of Gaussian distribution is difficult to verify in \ref{App1}, and discuss the behavior under an initial condition different from that of Section \ref{Simulation} in \ref{App2}.

\section{Model \label{Model} }

We consider the following one-dimensional Ising model with a long-range interaction:
\begin{equation}
{\cal H}  =  - J \sum _{i,j ; i \neq j } U \left( i-j \right) \sigma _{i} \sigma _{j}  , \label{Hamiltonian}
\end{equation}
\begin{equation}
\mathrm{where} \ \ \ \sigma _i =  \pm 1 . \label{Ising_spin}
\end{equation}
Here, we impose the periodic boundary condition and investigate the dynamics under the Markov chain Monte Carlo (MCMC) method. The coefficient $U(x)$ is given as
\begin{eqnarray}
U(x) & = & \frac{1}{N^{\ast} } \cdot \frac{1}{ |x| ^\alpha} \ \ \mathrm{if} \ \ |x| \leq \frac{N}{2} , \label{Ux} \\
U(x+N) & = & U(x) ,
\end{eqnarray}
where the constant $N^{\ast}$ is defined as 
\begin{equation}
 N^{\ast} = \sum _{ |x| \leq N/2, x \neq 0 } \frac{1}{|x|^\alpha} , \label{Nstar}
\end{equation}
so that the sum of $U(x)$ is normalized as 1:
\begin{equation}
\sum _{j \neq i} U(i-j) = 1 . \label{Unormal}
\end{equation}
Note that $U(x)$ is modified from the power function in accordance with the periodic boundary condition. In this paper, we consider the case in which $0 < \alpha \leq 1$. The order of $N^{\ast}$ can be calculated as follows: 
\begin{equation}
 N^{\ast} \sim \left\{ 
\begin{array}{c}
N^{1-\alpha} \ \ \mathrm{if} \ \ 0<\alpha<1 \\
\log N \ \ \mathrm{if} \ \ \alpha=1 \\
\end{array} 
\right. . \label{Nstar2}
\end{equation}
Considering that $N^{\ast } < N$, the small parameter of the perturbation is $O(1/N^{\ast} )$. 
The updating of each step is the flipping of one randomly chosen spin $\sigma _i$, that is,
\begin{equation}
\left\{ \sigma \right\} = (\sigma _1 , \sigma _2 , ... , \sigma _N ) \mapsto \left\{ F_i \sigma \right\} \equiv (\sigma _1 , \sigma _2 , ... , - \sigma _i , ... , \sigma _N ) ,
\end{equation}
and we define the unit of time $t$ as 1 Monte Carlo step (MCS). The time development of the probability distribution of the spin configuration, $P_N ( \left\{ \sigma \right\} ; t )$, is expressed as follows:
\begin{eqnarray}
 P_N \left( \left\{ \sigma \right\} ; t+ \frac{1}{N} \right) & = & P_N ( \left\{ \sigma \right\} ; t ) \nonumber \\
& & + \frac{1}{N} \sum _i \bigl\{ P_N ( \left\{ F_i \sigma \right\} ; t ) W \left( \left\{ F_i \sigma \right\} \rightarrow \left\{ \sigma \right\} \right) \bigr. \nonumber \\
& & \bigl. - P_N ( \left\{ \sigma \right\} ; t ) W( \left\{ \sigma \right\} \rightarrow \left\{ F_i \sigma \right\} ) \bigr\} ,
 \label{MF_dynam1}
\end{eqnarray}
where $W( \left\{ \sigma \right\} \rightarrow \left\{ \sigma ' \right\} )$ is the acceptance ratio of the updating, $ \left\{ \sigma \right\} \rightarrow \left\{ \sigma ' \right\} $. Note that $1/N$ MCS implies one step of updating. In this study, for simplicity, we considered a case in which the system is spatially homogeneous. In the typical MCMC method, $W$ is a function of the product of the energy change during updating and the inverse temperature.
\begin{equation}
\beta \delta E ( \left\{ \sigma \right\} \rightarrow \left\{ F_i \sigma \right\} ) = -2\beta J \cdot ( - \sigma _i - \sigma _i ) \sum _{ j \neq i } U(i-j) \sigma _j = 4 \beta J \sigma _i \sum _{ j \neq i } U(i-j) \sigma _j ; 
\end{equation}
therefore, we can express 
\begin{equation}
 W( \left\{ \sigma \right\} \rightarrow \left\{ F_i \sigma \right\} ) = w \left(  4 \beta J \sigma _i \sum _{ j \neq i } U(i-j) \sigma _j \right) .
 \label{transition0}
\end{equation}
There are several ways of defining the form of $w$. In this study, we adopted Glauber dynamics, defined as follows:
\begin{equation}
w( x ) = \frac{1}{1 + e^x } = \frac{1 - \tanh \left( \frac{x}{2} \right) }{2} .
\label{transition1}
\end{equation}
Assuming the relation $ \sum _{ j \neq i } U(i-j) \sigma _j \simeq m $, where $m$ is the ensemble average of the magnetization per one spin, we obtain the Taylor expansion of $W$ as
\begin{eqnarray}
 W( \left\{ \sigma \right\} \rightarrow \left\{ F_i \sigma \right\} ) & = & w \left(  4 \beta J m \sigma _i  \right) + 4 \beta J \sigma _i \sum _{ j \neq i } U(i-j) \left( \sigma _j -m \right) w' \left( 4 \beta J m \sigma _i  \right) \nonumber \\
& & + 8 \left( \beta J \right) ^2 \left[ \sum _{ j \neq i }  U(i-j) \left( \sigma _j -m \right) \right]^2 w'' \left( 4 \beta J m \sigma _i  \right) . \label{transition2}
\end{eqnarray}
In the right-hand side of this equation, the first term represents the value of $W$ in the mean field approximation, and the remaining two terms are the difference from this approximation.

\section{Calculations \label{Calculation} }

First, we define $P_s$ as
\begin{equation}
 P_{s, i_1 , i_2 , ... , i_s} ( \sigma _{i_1} , \sigma _{i_2} , ... , \sigma _{i_s} ; t ) \equiv \mathrm{Tr} _{ \left\{ \sigma _n \right\} _{n \neq i_1 , i_2 , ... , i_s } } P_N ( \left\{ \sigma \right\} ; t ) .
 \label{Ps0}
\end{equation}
In the case of the mean field approximation, which is the zeroth approximation of this study, the terms of (\ref{transition2}) containing the derivatives of $w$ are ignored. Under this approximation, the time development of $P_s$ under (\ref{MF_dynam1}) can be expressed as
\begin{equation}
 P_{s, i_1 , i_2 , ... , i_s} ( \sigma _{i_1} , \sigma _{i_2} , ... , \sigma _{i_s} ; t ) = \prod _{ n=1} ^s p_1 ( \sigma _{i_n} , t ) ,
 \label{Ps1}
\end{equation}
if $s$ is an $O(1)$ natural number. Here, $p_1 ( \sigma ; t )$ is the solution of the following equation:
\begin{eqnarray}
 p_1 \left( \sigma , t+ \frac{1}{N} \right) & = & p_1 ( \sigma , t ) \nonumber \\
& & + \frac{1}{N} \bigl\{ p_1 ( -\sigma , t ) w \left( - 4 \beta J m \sigma \right) - p_1 ( \sigma , t ) w \left(  4 \beta J m \sigma \right)  \bigr\} .
 \label{MF_dynam_1body}
\end{eqnarray}
Derivation of this equation is similar to the case of the infinite-range model. Considering that the relation between $p_1$ and the magnetization $m$ is expressed as
\begin{equation}
 p_1 ( \sigma , t ) = \frac{1+m \sigma }{2} ,
 \label{p1_m}
\end{equation}
(\ref{MF_dynam_1body}) can also be expressed as
\begin{equation}
 m \left( t+ \frac{1}{N} \right) = m ( t ) + \frac{1}{N} \bigl\{ (1-m) w \left( - 4 \beta J m \right) - (1+m) w \left(  4 \beta J m \right)  \bigr\} .
 \label{MF_dynam_m}
\end{equation}
Taking the thermodynamic limit, (\ref{MF_dynam_m}) is transformed into an ordinary differential equation \cite{PHB89,SK68,CA99,OYCK00}
\begin{eqnarray}
\frac{d m}{dt} & = & \bigl\{ (1-m) w \left( - 4 \beta J m \right) - (1+m) w \left(  4 \beta J m \right)  \bigr\} \nonumber \\
& = & -m + \tanh (2 \beta J m) .
 \label{MF_dynam_m2}
\end{eqnarray}

Taking the difference between the function $P_s$ and its zeroth approximation expressed in (\ref{Ps1}) as $\delta p_s$, we obtain the following expressions:
\numparts
\begin{eqnarray}
P_{1,i} ( \sigma _i , t ) & \equiv & p_1(\sigma _i , t) + \delta p_{1,i} ( \sigma _i , t ) , \label{P1} \\
P_{2,i,j} ( \sigma _i , \sigma _j ; t ) & \equiv & p_1(\sigma _i , t) p_1(\sigma _j , t) + \delta p_{2,i,j} ( \sigma _i , \sigma _j ; t ) , \label{P2} \\
P_{3,i,j,k} ( \sigma _i , \sigma _j , \sigma _k ; t ) & \equiv & p_1(\sigma _i , t) p_1(\sigma _j , t) p_1(\sigma _k , t) + \delta p_{3,i,j,k} ( \sigma _i , \sigma _j , \sigma _k ; t ) , \label{P3} \\
P_{4,i,j,k,l} ( \sigma _i , \sigma _j , \sigma _k , \sigma _l ; t ) & \equiv & p_1(\sigma _i , t) p_1(\sigma _j , t) p_1(\sigma _k , t) p_1(\sigma _l , t) + \delta p_{4,i,j,k,l} ( \sigma _i , \sigma _j , \sigma _k , \sigma _l ; t ) , \label{P4}
\end{eqnarray}
\endnumparts
\begin{eqnarray}
\mathrm{where} \ \ \ \delta p_{2,i,j} ( \sigma _i , \sigma _j ; t ) , \delta p_{3,i,j,k} ( \sigma _i , \sigma _j , \sigma _k ; t ) , \delta p_{4,i,j,k,l} ( \sigma _i , \sigma _j , \sigma _k , \sigma _l ; t ) = O \left( \frac{1}{N^{\ast}} \right) .
 \label{P2a}
\end{eqnarray}
In this paper, the characters $i,j,k,$ and $l$ represent different numbers, unless there is a summation over them, such as $\sum _{i,j}$. Note that $p_1$ is not defined as the probability distribution of a single spin in a finite-size system. Instead, it is that in the thermodynamic limit. Specifically, $p_1$ itself is defined as the solution of (\ref{MF_dynam_1body}), and the modification of the finite-size system appears as $\delta p_1$, $\delta p_2$, $\delta p_3$, and $\delta p_4$. We do not consider $P_s$ with $s \geq 5$ because it was not required for the calculation in this study.
Assuming that the magnetization does not depend on space, $\delta p_2$ is expressed as
\begin{eqnarray}
\delta p_{2,i,j} ( \sigma _i , \sigma _j ; t ) & = & \frac{\delta m}{4} \left( \sigma _i + \sigma _j \right) + \frac{v_{ij} + 2m \delta m }{4} \sigma _i \sigma _j . \label{p2} 
\end{eqnarray}
Here, $\delta m$ is the modification of (the ensemble average of) the magnetization.

To derive (\ref{p2}), we used the fact that arbitrary function $f(\sigma _i ,\sigma _j)$ of the two spin variables $\sigma _i$ and $\sigma _j$ has the form
expressed as
\begin{equation}
f(\sigma _i , \sigma _j) = \gamma _1 + \gamma _2 \sigma _i + \gamma _3 \sigma _j + \gamma _4 \sigma_i \sigma_j ,
\label{2body_general}
\end{equation}
where $\gamma_1, \gamma_2, \gamma_3$, and $\gamma_4$ are the constants determined by the values of $f$ as
\numparts
\begin{eqnarray}
\gamma _1 & = & \frac{1}{4} \sum_{ \sigma _i ,\sigma _j} f (\sigma _i , \sigma _j )  , \label{2body_general_2a} \\
\gamma _2 & = & \frac{1}{4} \sum_{ \sigma _i ,\sigma _j} \sigma _i f (\sigma _i , \sigma _j ) , \label{2body_general_2b} \\
\gamma _3 & = & \frac{1}{4} \sum_{ \sigma _i ,\sigma _j} \sigma _j f (\sigma _i , \sigma _j ) , \label{2body_general_2c} \\
\gamma _4 & = & \frac{1}{4} \sum_{ \sigma _i ,\sigma _j} \sigma _i \sigma _j f (\sigma _i , \sigma _j ) . 
\label{2body_general_2d}
\end{eqnarray}
\endnumparts
In the case of $\delta p _2$, the following conditions exist,
\numparts
\begin{eqnarray}
\sum_{ \sigma _i ,\sigma _j} \delta p_{2,i,j} (\sigma _i ,\sigma _j ;t) & = & \sum_{ \sigma _i ,\sigma _j} \left\{ P_{2,i,j} (\sigma _i ,\sigma _j ;t) - p_1 (\sigma _i , t ) p_1 (\sigma _j , t ) \right\} = 1-1 = 0 , \label{2body_gamma1} \\
\sum_{ \sigma _i ,\sigma _j} \sigma _i \delta p_{2,i,j} (\sigma _i ,\sigma _j ;t) & = & \sum_{ \sigma _i ,\sigma _j} \sigma _i  \left\{ P_{2,i,j} (\sigma _i ,\sigma _j ;t) - p_1 (\sigma _i , t ) p_1 (\sigma _j , t ) \right\} \nonumber \\ & =  & (m + \delta m) - m = \delta m ,  \label{2body_gamma2} \\
\sum_{ \sigma _i ,\sigma _j} \sigma _j \delta p_{2,i,j} (\sigma _i ,\sigma _j ;t) & = & \delta m , \label{2body_gamma3} 
\end{eqnarray}
\endnumparts
hence three of the constants that appeared in (\ref{2body_general}) are given as $\gamma _1 = 0$ and $\gamma_2 = \gamma_3 = \delta m/4$. The remaining constant, $\gamma _4$, cannot be determined by the normalization condition or the value of magnetization. Introducing the quantity $v_{ij}$ which fulfills $\gamma _4 = (v_{ij} + 2m \delta m)/4$, we obtain (\ref{p2}). Under this definition, $v_{ij}$ coincides with (the ensemble average of) the spatial correlation function if we ignore higher-order perturbation terms:
\begin{eqnarray}
\left< \sigma _i \sigma _j \right> - \left< \sigma _i \right> \left< \sigma _j \right> & = & \sum _{\sigma _i, \sigma _j} \sigma _i \sigma _j \left( p_1 (\sigma _i) p_1 (\sigma _j) + \delta p_{2,i,j} ( \sigma _i , \sigma _j ; t ) \right) - (m + \delta m)^2 \nonumber \\
& = & v_{ij} + O(\delta m ^2) . \label{vij} 
\end{eqnarray}
When the system is spatially homogeneous, $v_{ij}$ depends on $i$ and $j$ only through the distance between them, $\left| i- j \right|$.
Note that the probability distribution of two spin variables $P_2$ are expressed by macroscopic quantities such as the magnetization and the spatial correlation through (\ref{p1_m}) and (\ref{p2}). Considering this point, calculation of the quantities $m$, $\delta m$, and $v_{ij}$ is equivalent to that of the probability distribution $P_2$ itself. This discussion is based on the point that each variable of Ising model can have only two values, $\pm 1$. In the case of the more complicated models, the simple relation such as (\ref{2body_general}) do not exist and the form of $P_2$ will be more difficult.


As shown later, the equation describing the time development of $P_2$ does not have a closed form because it includes $\delta p_3$ and $\delta p_4$. This is a reflection of BBGKY hierarchy \cite{CDFR14}. In this study, we assumed that the probability density of any normalized extensive properties expressed in the form $X/N = \sum _i x(\sigma _i)/N $ approximately obey a Gaussian form. Here, $x(\sigma _i)$ is an arbitrary function of $\sigma _i$. Hence, the third- and fourth-order cumulants of $X/N$ are nearly zero.
\begin{equation}
\frac{1}{N^3} \left( \left< \sum _{i,j,k} x(\sigma _i) x(\sigma _j) x(\sigma _k) \right> - 3 \left< \sum _{i,j} x(\sigma _i) x(\sigma _j) \right> \left< \sum _i x(\sigma _i) \right> + 2 \left< \sum _i x(\sigma _i) \right> ^3 \right) = 0 . 
 \label{mv2a}
\end{equation}
\begin{equation}
\frac{1}{N^4} \left( \left< \sum _{i,j,k,l} x(\sigma _i) x(\sigma _j) x(\sigma _k) x(\sigma _l) \right> - 3 \left< \sum _{i,j} x(\sigma _i) x(\sigma _j) \right> ^2 + 2 \left< \sum _i x(\sigma _i) \right> ^4 \right) = 0 . 
 \label{mv2a_4th}
\end{equation}
This assumption is a generalization of a similar assumption from our previous study, which demands a Gaussian distribution of magnetization. In the case of the previous study, we could verify the validity of this assumption at least in an equilibrium state. However, it is difficult to generalize this discussion to the case of our present model, hence comparison between the numerical simulation discussed in the next section is important. We explain why this generalization is difficult in \ref{App1}.

Using (\ref{P2}) and (\ref{P3}), (\ref{mv2a}) can be transformed as
\begin{eqnarray}
& & \frac{1}{N^3} \sum _{i,j,k} \sum _{\sigma_i , \sigma_j , \sigma_k} x(\sigma _i) x(\sigma _j) x(\sigma _k) \left\{ \delta p_3 ( \sigma _i , \sigma _j , \sigma _k ; t ) \right. \nonumber \\
& & \left. - 3 \delta p_2 ( \sigma _i , \sigma _j ; t ) p_1 ( \sigma _k , t ) + 3 \delta p_1 ( \sigma _i , t ) p_1 ( \sigma _j , t ) p_1 ( \sigma _k , t ) + O \left( \frac{1}{N^{\ast 2}} \right) \right\} = 0 . \nonumber \\
 \label{mv2b}
\end{eqnarray}
To derive this relation, we used the fact that the $O \left( 1 \right)$ terms of (\ref{mv2a}) cancel each other out.
If $\delta p _3 $ obeys 
\begin{eqnarray}
\delta p_{3,i,j,k} ( \sigma _i , \sigma _j , \sigma _k ; t ) & = & \delta p_{2,i,j} ( \sigma _i , \sigma _j ; t ) p_{1,k} (\sigma _k , t) + \delta p_{2,j,k} ( \sigma _j , \sigma _k ; t ) p_{1,i} (\sigma _i , t) \nonumber \\
& &  + \delta p_{2,k,i} ( \sigma _k , \sigma _i ; t ) p_{1,j} (\sigma _j , t) - \delta p _{1,i} (\sigma _i , t) p _{1,j} (\sigma _j , t) p _{1,k} (\sigma _k , t) \nonumber \\
& &  - \delta p _{1,j} (\sigma _j , t) p _{1,k} (\sigma _k , t) p _{1,i} (\sigma _i , t) - \delta p _{1,k} (\sigma _k , t) p _{1,i} (\sigma _i , t) p _{1,j} (\sigma _j , t) \nonumber \\
& & + O \left( \frac{1}{N^{\ast 2}} \right) , 
 \label{alpha3}
\end{eqnarray} 
(\ref{mv2b}) holds for arbitrary $x$. Note that $\delta p_3$ is invariant under the exchanging of the suffixes $i,j$, and $k$. Ignoring the $O \left( 1/N^{\ast 2} \right)$ terms, (\ref{alpha3}) is equivalent to the Kirkwood superposition approximation \cite{Kirkwood42,CL64}.
\begin{eqnarray}
P_{3,i,j,k} ( \sigma _i , \sigma _j , \sigma _k ; t ) & = & \frac{P_{2,i.j} ( \sigma _i , \sigma _j ; t ) P_{2,j,k} ( \sigma _j , \sigma _k ; t ) P_{2,k,i} ( \sigma _k , \sigma _i ; t )}{ P_{1,i} (\sigma _i , t) P_{1,j} (\sigma _j , t) P_{1,k} (\sigma _k , t)} . 
 \label{Kirkwood3}
\end{eqnarray}
This equivalence also exists in the case of the infinite-range model treated in our previous study, although we did not perceive it.
Similar calculations with (\ref{mv2a_4th}) give the expression for $\delta p_4$,
\begin{eqnarray}
\delta p_{4,i,j,k,l} ( \sigma _i , \sigma _j , \sigma _k , \sigma _l ; t ) & = & \delta p_{2,i,j} ( \sigma _i , \sigma _j ; t ) p_{1,k} (\sigma _k , t) p_{1,l} (\sigma _l , t) + \delta p_{2,i,k} ( \sigma _i , \sigma _k ; t ) p_{1,j} (\sigma _j , t) p_{1,l} (\sigma _l , t) \nonumber \\
& & + \delta p_{2,i,l} ( \sigma _i , \sigma _l ; t ) p_{1,j} (\sigma _j , t) p_{1,k} (\sigma _k , t) + \delta p_{2,j,k} ( \sigma _j , \sigma _k ; t ) p_{1,i} (\sigma _i , t) p_{1,l} (\sigma _l , t) \nonumber \\
& & + \delta p_{2,j,l} ( \sigma _j , \sigma _l ; t ) p_{1,i} (\sigma _i , t) p_{1,k} (\sigma _k , t) + \delta p_{2,k,l} ( \sigma _k , \sigma _l ; t ) p_{1,i} (\sigma _i , t) p_{1,j} (\sigma _j , t) \nonumber \\
& & - 2 \delta p _{1,i} (\sigma _i , t) p _{1,j} (\sigma _j , t) p _{1,k} (\sigma _k , t) p_{1,l} (\sigma _l , t) \nonumber \\
& & - 2 \delta p _{1,j} (\sigma _j , t) p _{1,k} (\sigma _k , t) p_{1,l} (\sigma _l , t) p _{1,i} (\sigma _i , t) \nonumber \\
& & - 2 \delta p _{1,k} (\sigma _k , t) p_{1,l} (\sigma _l , t) p _{1,i} (\sigma _i , t) p _{1,j} (\sigma _j , t) \nonumber \\
& & - 2 \delta p _{1,l} (\sigma _l , t) p _{1,i} (\sigma_i , t) p _{1,j} (\sigma _j , t) p_{1,k} (\sigma _k , t) + O \left( \frac{1}{N^{\ast 2}} \right) , 
 \label{alpha4}
\end{eqnarray} 
which is equivalent to the following equation, ignoring the $O \left( 1/N^{\ast 2} \right)$ terms:
\begin{eqnarray}
& & P_{4,i,j,k,l} ( \sigma _i , \sigma _j , \sigma _k , \sigma _l ; t ) \nonumber \\
& = & \frac{P_{2,i,j} ( \sigma _i , \sigma _j ; t ) P_{2,i,k} ( \sigma _i , \sigma _k ; t ) P_{2,i,l} ( \sigma _i , \sigma _l ; t ) P_{2,j,k} ( \sigma _j , \sigma _k ; t ) P_{2,j,l} ( \sigma _j , \sigma _l ; t ) P_{2,k,l} ( \sigma _k , \sigma _l ; t )}{ \left( P_{1,i} (\sigma _i , t) P_{1,j} (\sigma _j , t) P_{1,k} (\sigma _k , t) P_{1,l} (\sigma _l , t) \right) ^2 } . 
 \label{Kirkwood4}
\end{eqnarray}
This relation is the generalization of the Kirkwood superposition approximation \cite{AJM97}.

\subsection{Calculation of the spin correlation \label{2body} }

Substituting (\ref{transition2}) into (\ref{MF_dynam1}), the master equation in the finite-$N$ case becomes
\begin{eqnarray}
P_N \left( \left\{ \sigma \right\} ; t+ \frac{1}{N} \right) & = & P_N ( \left\{ \sigma \right\} ; t ) \nonumber \\
& & + \frac{1}{N} \sum _i \left[ P_N ( \left\{ F_i \sigma \right\} ; t ) \biggl\{ w \left( - 4 \beta J m \sigma _i  \right) \biggr. \right. \nonumber \\
& & - 4 \beta J \sigma _i \sum _{j \neq i} U \left( i-j \right) \left( \sigma _j -m \right) w' \left( - 4 \beta J m \sigma _i  \right) \nonumber \\
& & \left. + 8 \left( \beta J \right) ^2 \left[ \sum _{j \neq i} U \left( i-j \right) \left( \sigma _j -m \right) \right] ^2 w'' \left( - 4 \beta J m \sigma _i  \right) \right\} \nonumber \\
& & - P_N ( \left\{ \sigma \right\} ; t ) \biggl\{ w \left(  4 \beta J m \sigma _i  \right) \biggr. \nonumber \\
& & + 4 \beta J \sigma _i \sum _{j \neq i} U \left( i-j \right) \left( \sigma _j -m \right) w' \left( 4 \beta J m \sigma _i  \right) \nonumber \\
& & \left. \left. + 8 \left( \beta J \right) ^2 \left[ \sum _{j \neq i} U \left( i-j \right) \left( \sigma _j -m \right) \right] ^2 w'' \left( 4 \beta J m \sigma _i  \right) \right\} \right] .
 \label{MF_dynam2_a}
\end{eqnarray}
Here, we consider the time development of this equation, starting from the zeroth-order approximation given by (\ref{Ps1}).

When the summation is taken over all spin variables except $\sigma _i$ and $\sigma _j$, (\ref{MF_dynam2_a}) can be transformed into
\begin{eqnarray}
& & P_{2,i,j} \left( \sigma _i , \sigma _j ; t+ \frac{1}{N} \right) - P_{2,i,j} ( \sigma _i , \sigma _j ; t ) \nonumber \\
& = & \frac{1}{N} \left[  w \left( - 4 \beta J m \sigma _i  \right) P_{2,i,j} ( - \sigma _i , \sigma _j ; t ) \right. \nonumber \\
& & -  \mathrm{Tr} _{ \left\{ \sigma _n \right\} _{n \neq i,j} } 4 \beta J \sigma _i \sum _{k\neq i} U \left( k-i \right) \left( \sigma _k - m \right) w' \left( - 4 \beta J m \sigma _i  \right) P_N ( \left\{ F_i \sigma \right\} ; t ) \nonumber \\
& & + \mathrm{Tr} _{ \left\{ \sigma _n \right\} _{n \neq i,j} } 8 \left( \beta J \right) ^2 \sum _{k,l \neq i} U \left( k-i \right) U \left( l-i \right) \left( \sigma _k - m \right) \left( \sigma _l - m \right) w'' \left( - 4 \beta J m \sigma _i  \right) P_N ( \left\{ F_i \sigma \right\} ; t ) \nonumber \\
& & + w \left( - 4 \beta J m \sigma _j  \right) P_{2,i,j} ( \sigma _i , - \sigma _j ; t ) \nonumber \\
& & - \mathrm{Tr} _{ \left\{ \sigma _n \right\} _{n \neq i,j} } 4 \beta J \sigma _j \sum _{k\neq j} U \left( k-j \right) \left( \sigma _k - m \right) w' \left( - 4 \beta J m \sigma _j  \right) P_N ( \left\{ F_j \sigma \right\} ; t ) \nonumber \\
& & + \mathrm{Tr} _{ \left\{ \sigma _n \right\} _{n \neq i,j} } 8 \left( \beta J \right) ^2 \sum _{k,l \neq j} U \left( k-j \right) U \left( l-j \right) \left( \sigma _k - m \right) \left( \sigma _l - m \right) w'' \left( - 4 \beta J m \sigma _j  \right) P_N ( \left\{ F_j \sigma \right\} ; t ) \nonumber \\
& & - w \left(  4 \beta J m \sigma _i  \right) P_{2,i,j} (  \sigma _i , \sigma _j ; t ) \nonumber \\
& & - \mathrm{Tr} _{ \left\{ \sigma _n \right\} _{n \neq i,j} } 4 \beta J \sigma _i \sum _{k\neq i} U \left( k-i \right) \left( \sigma _k - m \right) w' \left( 4 \beta J m \sigma _i  \right) P_N ( \left\{ \sigma \right\} ; t ) \nonumber \\ 
& & - \mathrm{Tr} _{ \left\{ \sigma _n \right\} _{n \neq i,j} } 8 \left( \beta J \right) ^2 \sum _{k,l \neq i} U \left( k-i \right) U \left( l-i \right) \left( \sigma _k - m \right) \left( \sigma _l - m \right) w'' \left( 4 \beta J m \sigma _i  \right) P_N ( \left\{ \sigma \right\} ; t ) \nonumber \\
& & -w \left(  4 \beta J m \sigma _j  \right) P_{2,i,j} (  \sigma _i , \sigma _j ; t ) \nonumber \\
& & - \mathrm{Tr} _{ \left\{ \sigma _n \right\} _{n \neq i,j} } 4 \beta J \sigma _j \sum _{k\neq j} U \left( k-j \right) \left( \sigma _k - m \right) w' \left( 4 \beta J m \sigma _j  \right) P_N ( \left\{ \sigma \right\} ; t ) \nonumber \\
& & \left. - \mathrm{Tr} _{ \left\{ \sigma _n \right\} _{n \neq i,j} } 8 \left( \beta J \right) ^2 \sum _{k,l \neq j} U \left( k-j \right) U \left( l-j \right) \left( \sigma _k - m \right) \left( \sigma _l - m \right) w'' \left( 4 \beta J m \sigma _j  \right) P_N ( \left\{ \sigma \right\} ; t ) \right] . \nonumber \\
 \label{MF_dynam2_b}
\end{eqnarray}
To calculate the perturbation, we should first note that some terms appearing in the infinite-range model of our previous study become the higher-order infinitesimals \cite{Komatsu22}, because of the difference of the small parameter or the shape of the interaction. One example is seen in the left-hand side of (\ref{MF_dynam2_b}), which is expressed as
\begin{eqnarray}
& & \left( \frac{1}{N} \right) ^{-1} \cdot \left[ P_2 \left( \sigma _i , \sigma _j ; t+ \frac{1}{N} \right) - P_2 ( \sigma _i , \sigma _j ; t ) \right] \nonumber \\
& = & \left( \frac{1}{N} \right) ^{-1} \cdot \left[ \left\{ p_1 \left( \sigma _i , t+ \frac{1}{N} \right) - p_1 \left( \sigma _i , t \right) \right\} p_1 \left( \sigma _j , t \right) \right. \nonumber \\ 
& & + p_1 \left( \sigma _i , t \right) \left\{ p_1 \left( \sigma _j , t+ \frac{1}{N} \right) - p_1 \left( \sigma _j , t \right) \right\} \nonumber \\
& & + \delta p _{2,i,j} \left( \sigma _i , \sigma _j ; t+ \frac{1}{N} \right) - \delta p _{2,i,j} \left( \sigma _i , \sigma _j ; t \right) \nonumber \\
& & \left. + \left\{ p_1 \left( \sigma _i , t+ \frac{1}{N} \right) - p_1 \left( \sigma _i , t \right) \right\} \left\{ p_1 \left( \sigma _j , t+ \frac{1}{N} \right) - p_1 \left( \sigma _j , t \right) \right\} \right] .
 \label{MF_dynam2_e}
\end{eqnarray}
The fourth term of the right-hand side of (\ref{MF_dynam2_e}) is $O(1/N)$, which is the higher-order infinitesimal in this study:
\begin{eqnarray}
& & \left( \frac{1}{N} \right) ^{-1} \left\{ p_1 \left( \sigma _i , t+ \frac{1}{N} \right) - p_1 \left( \sigma _i , t \right) \right\} \left\{ p_1 \left( \sigma _j , t+ \frac{1}{N} \right) - p_1 \left( \sigma _j , t \right) \right\} \nonumber \\
& = & O \left( \frac{1}{N} \right) = o \left( \frac{1}{N^{\ast} } \right) .
 \label{MF_higher_inf1}
\end{eqnarray}
This term is not negligible in the case of the infinite-range model, because the small parameter of this case is $1/N$ itself. 

To simplify the right-hand side of (\ref{MF_dynam2_b}), we calculate the traces in the bracket. For example, the second term of the above equation is transformed into 
\begin{eqnarray}
& & \mathrm{Tr} _{ \left\{ \sigma _n \right\} _{n \neq i,j} } 4 \beta J \sigma _i \sum _{k\neq i} U \left( k-i \right) \left( \sigma _k - m \right) w' \left( - 4 \beta J m \sigma _i  \right) P_N ( \left\{ F_i \sigma \right\} ; t ) \nonumber \\ 
& = & 4 \beta J \sigma _i \sum _{k \neq i,j} \sum _{ \sigma _k } U \left( k-i \right) \left( \sigma _k - m \right) w' \left( - 4 \beta J m \sigma _i  \right) P_{3,i,j,k} ( -\sigma _i , \sigma _j , \sigma _k ; t ) \nonumber \\
& & + 4 \beta J \sigma _i U \left( j-i \right) \left( \sigma _j - m \right) w' \left( - 4 \beta J m \sigma _i  \right) P_{2,i,j} ( -\sigma _i , \sigma _j ; t ) \nonumber \\
& = & 4 \beta J \sigma _i \sum _{k \neq i,j} \sum _{ \sigma _k } U \left( k-i \right) \left( \sigma _k - m \right)  w' \left( - 4 \beta J m \sigma _i  \right) \delta p_{3,i,j,k} ( -\sigma _i , \sigma _j , \sigma _k ; t ) \nonumber \\
& & + 4 \beta J \sigma _i U \left( j-i \right) \left( \sigma _j - m \right) w' \left( - 4 \beta J m \sigma _i  \right) p_1 ( -\sigma _i , t ) p_1( \sigma _j , t ) + o \left( \frac{1}{N^{\ast} } \right) .
 \label{approx_p3}
\end{eqnarray}
In the final line of (\ref{approx_p3}), an expansion of $P_3$, given as (\ref{P3}), and the relation
\begin{equation}
\sum _{ \sigma _k } \sigma _k p_1(\sigma _k , t) = m  \label{s_average}
\end{equation}
are used. Substituting (\ref{alpha3}), the first term of the right-hand side of (\ref{approx_p3}) is transformed as
\begin{eqnarray}
& & 4 \beta J \sigma _i \sum _{k \neq i,j} \sum _{ \sigma _k } U \left( k-i \right) \left( \sigma _k - m \right)  w' \left( - 4 \beta J m \sigma _i  \right) \delta p_{3,i,j,k} ( -\sigma _i , \sigma _j , \sigma _k ; t ) \nonumber \\
& = & 4 \beta J \sigma _i \sum _{k \neq i,j} \sum _{ \sigma _k } U \left( k-i \right) \left( \sigma _k - m \right)  w' \left( - 4 \beta J m \sigma _i  \right) \nonumber \\
& & \cdot \left[ \delta p_{2,i,k} ( -\sigma _i , \sigma _k ; t ) p_1( \sigma _j , t ) + \delta p_{2,j,k} ( \sigma _j , \sigma _k ; t ) p_1( - \sigma _i , t ) - \delta p_{1,k} ( \sigma _k , t ) p_1( - \sigma _i , t ) p_1(  \sigma _j , t ) \right] \nonumber \\
& = & 4 \beta J \sigma _i \sum _{k \neq i,j} U \left( k-i \right) w' \left( - 4 \beta J m \sigma _i  \right) \cdot \left[ \left( \frac{\delta m }{2} - \frac{ v_{ik} + m \delta m }{2} \sigma _i \right) p_1( \sigma _j , t ) \right. \nonumber \\
& & \left. + \left( \frac{\delta m }{2} + \frac{ v_{jk} + m \delta m }{2} \sigma _j \right) p_1( - \sigma _i , t ) - \delta m p_1( - \sigma _i , t ) p_1(  \sigma _j , t ) \right] \nonumber \\
& = & \beta J \sigma _i \sum _{k \neq i,j} U \left( k-i \right) w' \left( - 4 \beta J m \sigma _i  \right) \cdot \left[ \delta m \right. \nonumber \\
& &  \left .- (v_{ik} + m \delta m ) \sigma _i + (v_{jk} + m \delta m ) \sigma _j - \left( m v_{ik} + m v_{jk} + m ^2 \delta m \right) \sigma _i \sigma _j \right] .
 \label{approx_p3_2}
\end{eqnarray}
Here, the following relations are used:
\begin{eqnarray}
\sum _{ \sigma _k } \left( \sigma _k - m \right) \delta p_{1,k} ( \sigma _k , t ) & = & \delta m , \label{Trp1} \\ 
\sum _{ \sigma _k } \left( \sigma _k - m \right) \delta p_{2,i,k} ( \sigma _i , \sigma _k , t ) & = & \frac{\delta m }{2} + \frac{ v_{ik} + m \delta m }{2} \sigma _i . \label{Trp2}
\end{eqnarray}
Similarly, the third term of (\ref{MF_dynam2_b}), which is proportional to $w'' \left( - 4 \beta J m \sigma _i  \right) $, can be transformed into
\begin{eqnarray}
& & \mathrm{Tr} _{ \left\{ \sigma _n \right\} _{n \neq i,j} } 8 \left( \beta J \right) ^2 \sum _{k,l \neq i} U \left( k-i \right) U \left( l-i \right) \left( \sigma _k - m \right) \left( \sigma _l - m \right) w'' \left( - 4 \beta J m \sigma _i  \right) P_N ( \left\{ F_i \sigma \right\} ; t ) \nonumber \\ 
& = & 8 \left( \beta J \right) ^2 \sum _{k \neq i,j } \sum _{\sigma _k } \left\{ U \left( k-i \right) \left( \sigma _k - m \right) \right\} ^2 w'' \left( - 4 \beta J m \sigma _i  \right) p_1( -\sigma _i , t) p_1(\sigma _j , t) p_1(\sigma _k , t) \nonumber \\
& & + 8 \left( \beta J \right) ^2 \sum _{k,l \neq i} \sum _{\sigma_k , \sigma_l } U \left( k-i \right) U \left( l-i \right) \left( \sigma _k - m \right) \left( \sigma _l - m \right) \nonumber \\
& & \cdot w'' \left( - 4 \beta J m \sigma _i  \right) \delta p_{4,i,j,k,l} ( -\sigma _i , \sigma _j , \sigma _k , \sigma _l ; t )  + o \left( \frac{1}{N^{\ast} } \right) \nonumber \\
& = & 8 \left( \beta J \right) ^2 \sum _{k,l \neq i} \sum _{\sigma_k , \sigma_l } U \left( k-i \right) U \left( l-i \right) \left( \sigma _k - m \right) \left( \sigma _l - m \right) \nonumber \\
& & \cdot w'' \left( - 4 \beta J m \sigma _i  \right) \delta p_{4,i,j,k,l} ( -\sigma _i , \sigma _j , \sigma _k , \sigma _l ; t )  + o \left( \frac{1}{N^{\ast} } \right) .
 \label{approx_p4}
\end{eqnarray}
The term containing $U(k-i) ^2$ can be ignored because 
\begin{equation}
\sum _{k \neq i } U \left( k-i \right)  ^2 = \frac{1}{N^{\ast} } \cdot \left( \frac{1}{N^{\ast} } \sum _{k \neq i } \frac{1}{| k-i |^{2\alpha} } \right) = o \left( \frac{1}{N^{\ast} } \right) .
\end{equation}
This term does not vanish in the case of the infinite-range model previously considered \cite{Komatsu22}. Substituting (\ref{alpha4}), the right-hand side of (\ref{approx_p4}) can be transformed as
\begin{eqnarray}
& & 8 \left( \beta J \right) ^2 \sum _{k,l \neq i} \sum _{\sigma_k , \sigma_l }  U \left( k-i \right) U \left( l-i \right) \left( \sigma _k - m \right) \left( \sigma _l - m \right) \nonumber \\
& & \cdot w'' \left( - 4 \beta J m \sigma _i  \right) \delta p_{2,k,l} ( \sigma _k , \sigma _l ; t ) p_1 ( -\sigma _i , t ) p_1 ( \sigma _j , t ) + o \left( \frac{1}{N^{\ast} } \right) \nonumber \\
& = & 8 \left( \beta J \right) ^2 \sum _{k,l \neq i}  U \left( k-i \right) U \left( l-i \right) w'' \left( - 4 \beta J m \sigma _i  \right)  v_{kl} p_1 ( -\sigma _i , t ) p_1 ( \sigma _j , t ) + o \left( \frac{1}{N^{\ast} } \right) \nonumber \\
& = & 8 \left( \beta J \right) ^2 v_{\infty} w'' \left( - 4 \beta J m \sigma _i  \right) p_1 ( -\sigma _i , t ) p_1 ( \sigma _j , t ) + o \left( \frac{1}{N^{\ast} } \right) ,
 \label{approx_p4_2}
\end{eqnarray}

\begin{equation}
\mathrm{where} \ \ v_{\infty} = \lim _{\left| k-l \right| \rightarrow \infty } v_{kl} .
\end{equation}
In the final line of (\ref{approx_p4_2}), we ignore the contribution of $\left( v_{kl} - v_{\infty}\right)$, which is the part of $v_{kl}$ that converges to zero at $\left| k-l \right| \rightarrow \infty$, because it will become the higher-order infinitesimal
\begin{equation}
\sum _{k,l \neq i}  U \left( k-i \right) U \left( l-i \right) \left( v_{kl} - v_{\infty}\right) = o \left( \frac{1}{N^{\ast} } \right) .
\end{equation}
Note that (\ref{approx_p4_2}) has a simpler form than (\ref{approx_p3_2}), although the Kirkwood superposition approximation of $\delta p_4$ (that is, (\ref{alpha4})) contains more terms than that of $\delta p_3$ (that is, (\ref{alpha3})). This is because most of the terms appearing in (\ref{alpha4}) vanish after the summation over $\sigma _k$ and $\sigma _l$, considering relation (\ref{s_average}). If we consider the higher-order terms of the Taylor expansion of $W$ omitted from (\ref{transition2}), we must calculate similar summations for $\delta p _s$ with $s \geq 5$. However, these summations completely vanish as a result of (\ref{s_average}). Hence, the contribution of these higher-order terms can be ignored.

 Substituting these equations into (\ref{MF_dynam2_b}), we obtain
\begin{eqnarray}
& & \left( \frac{1}{N} \right) ^{-1} \cdot \left[ P_2 \left( \sigma _i , \sigma _j ; t+ \frac{1}{N} \right) - P_2 ( \sigma _i , \sigma _j ; t ) \right] \nonumber \\
& = & w \left( - 4 \beta J m \sigma _i  \right) P_{2,i,j} ( - \sigma _i , \sigma _j ; t ) + w \left( - 4 \beta J m \sigma _j  \right) P_{2,i,j} ( \sigma _i , - \sigma _j ; t ) \nonumber \\
& & - w \left(  4 \beta J m \sigma _i  \right) P_{2,i,j} (  \sigma _i , \sigma _j ; t ) - w \left(  4 \beta J m \sigma _j  \right) P_{2,i,j} (  \sigma _i , \sigma _j ; t ) \nonumber \\
& & - 4 \beta J \sigma _i U \left( j-i \right) \left( \sigma _j - m \right) w' \left( - 4 \beta J m \sigma _i  \right) p_1 ( -\sigma _i , t ) p_1( \sigma _j , t ) \nonumber \\
& & - 4 \beta J \sigma _j U \left( j-i \right) \left( \sigma _i - m \right) w' \left( - 4 \beta J m \sigma _j  \right) p_1 ( \sigma _i , t ) p_1( -\sigma _j , t ) \nonumber \\
& & - 4 \beta J \sigma _i U \left( j-i \right) \left( \sigma _j - m \right) w' \left( 4 \beta J m \sigma _i  \right) p_1 ( \sigma _i , t ) p_1( \sigma _j , t ) \nonumber \\
& & - 4 \beta J \sigma _j U \left( j-i \right) \left( \sigma _i - m \right) w' \left( 4 \beta J m \sigma _j  \right) p_1 ( \sigma _i , t ) p_1( \sigma _j , t ) \nonumber \\
& & - 4 \beta J \sigma _i \sum _{k \neq i,j} \sum _{ \sigma _k } U \left( k-i \right) \left( \sigma _k - m \right)  w' \left( - 4 \beta J m \sigma _i  \right) \delta p_{3,i,j,k} ( -\sigma _i , \sigma _j , \sigma _k ; t ) \nonumber \\
& & - 4 \beta J \sigma _j \sum _{k \neq i,j} \sum _{ \sigma _k } U \left( k-j \right) \left( \sigma _k - m \right)  w' \left( - 4 \beta J m \sigma _j  \right) \delta p_{3,i,j,k} ( \sigma _i , -\sigma _j , \sigma _k ; t ) \nonumber \\
& & - 4 \beta J \sigma _i \sum _{k \neq i,j} \sum _{ \sigma _k } U \left( k-i \right) \left( \sigma _k - m \right)  w' \left( 4 \beta J m \sigma _i  \right) \delta p_{3,i,j,k} ( \sigma _i , \sigma _j , \sigma _k ; t ) \nonumber \\ 
& & - 4 \beta J \sigma _j \sum _{k \neq i,j} \sum _{ \sigma _k } U \left( k-j \right) \left( \sigma _k - m \right)  w' \left( 4 \beta J m \sigma _j  \right) \delta p_{3,i,j,k} ( \sigma _i , \sigma _j , \sigma _k ; t ) \nonumber \\
& & + 8 \left( \beta J \right) ^2 v_{\infty} w'' \left( - 4 \beta J m \sigma _i  \right) p_1 ( -\sigma _i , t ) p_1 ( \sigma _j , t ) \nonumber \\
& & + 8 \left( \beta J \right) ^2 v_{\infty} w'' \left( - 4 \beta J m \sigma _j  \right) p_1 ( \sigma _i , t ) p_1 ( -\sigma _j , t ) \nonumber \\
& & - 8 \left( \beta J \right) ^2 v_{\infty} w'' \left( 4 \beta J m \sigma _i  \right) p_1 ( \sigma _i , t ) p_1 ( \sigma _j , t ) \nonumber \\
& & - 8 \left( \beta J \right) ^2 v_{\infty} w'' \left( 4 \beta J m \sigma _j  \right) p_1 ( \sigma _i , t ) p_1 ( \sigma _j , t ) + o \left( 1/N ^{\ast} \right) \nonumber \\ 
& = & w \left( - 4 \beta J m \sigma _i  \right) P_{2,i,j} ( - \sigma _i , \sigma _j ; t ) + w \left( - 4 \beta J m \sigma _j  \right) P_{2,i,j} ( \sigma _i , - \sigma _j ; t ) \nonumber \\
& & - w \left(  4 \beta J m \sigma _i  \right) P_{2,i,j} (  \sigma _i , \sigma _j ; t ) - w \left(  4 \beta J m \sigma _j  \right) P_{2,i,j} (  \sigma _i , \sigma _j ; t ) \nonumber \\
& & - 4 \beta J U \left( j-i \right) w' \left( 4 \beta J m \right) \left( 1- m ^2 \right) \sigma _i \sigma _j \nonumber \\
& & - 2 \beta J w' \left( 4 \beta J m \right) \left\{ \delta m ( \sigma _i + \sigma _j ) + 2 ( v _{\infty } + m \delta m ) \sigma _i \sigma _j \right\} \nonumber \\
& & - 4 \left( \beta J \right) ^2 v_{\infty} w'' \left( 4 \beta J m \right) \left( \sigma _i + \sigma _j + 2m \sigma _i \sigma _j \right) + o \left( 1/N ^{\ast} \right) .
 \label{MF_dynam2_d}
\end{eqnarray}
In the final transformation of (\ref{MF_dynam2_d}), we use the fact that $w'(x)$ is an even function and $w''(x)$ is an odd function. The $O(1)$ terms of (\ref{MF_dynam2_d}) cancel each other out. Taking the limit $N \rightarrow \infty$ for the remaining terms, we can obtain an equation describing the time development of $\delta p_2$.
\begin{eqnarray}
& & \frac{d}{dt} \delta p _{2,i,j} \left( \sigma _i , \sigma _j ; t \right) \nonumber \\
& = & w \left( - 4 \beta J m \sigma _i  \right) \delta p _{2,i,j} ( - \sigma _i , \sigma _j ; t ) + w \left( - 4 \beta J m \sigma _j  \right) \delta p _{2,i,j} ( \sigma _i , - \sigma _j ; t ) \nonumber \\
& & - w \left(  4 \beta J m \sigma _i  \right) \delta p _{2,i,j} (  \sigma _i , \sigma _j ; t ) - w \left(  4 \beta J m \sigma _j  \right) \delta p_{2,i,j} (  \sigma _i , \sigma _j ; t ) \nonumber \\
& & - 4 \beta J U \left( j-i \right) w' \left( 4 \beta J m \right) \left( 1- m ^2 \right) \sigma _i \sigma _j \nonumber \\
& & - 2 \beta J w' \left( 4 \beta J m \right) \left\{ \delta m ( \sigma _i + \sigma _j ) + 2 ( v _{\infty } + m \delta m ) \sigma _i \sigma _j \right\} \nonumber \\
& & - 4 \left( \beta J \right) ^2 v_{\infty} w'' \left( 4 \beta J m \right) \left( \sigma _i + \sigma _j + 2m \sigma _i \sigma _j \right) \nonumber \\
& = & - \frac{\delta m }{4} \left( \sigma _i + \sigma _j  \right) \bigl\{ w \left( - 4 \beta J m \right) + w \left( 4 \beta J m \right) \bigr\} + \frac{\delta m }{2} \sigma _i \sigma _j \bigl\{ w \left( - 4 \beta J m \right) - w \left( 4 \beta J m \right) \bigr\} \nonumber \\
& & - \frac{v _{ij} + 2m \delta m}{2} \sigma _i \sigma _j \bigl\{ w \left( - 4 \beta J m \right) + w \left( 4 \beta J m \right) \bigr\} \nonumber \\
& & - 4 \beta J U \left( j-i \right) w' \left( 4 \beta J m \right) \left( 1- m ^2 \right) \sigma _i \sigma _j \nonumber \\
& & - 2 \beta J w' \left( 4 \beta J m \right) \left\{ \delta m ( \sigma _i + \sigma _j ) + 2 ( v _{\infty } + m \delta m ) \sigma _i \sigma _j \right\} \nonumber \\
& & - 4 \left( \beta J \right) ^2 v_{\infty} w'' \left( 4 \beta J m \right) \left( \sigma _i + \sigma _j + 2m \sigma _i \sigma _j \right) \nonumber \\
\label{MF_dynam2body_a}
\end{eqnarray}

Comparing (\ref{p2}) and (\ref{MF_dynam2body_a}), we obtain the following equations:
\begin{eqnarray}
& & \frac{1}{4} \frac{d}{dt} \delta m \nonumber \\
& = & - \frac{\delta m }{4} \bigl\{ w \left( - 4 \beta J m \right) + w \left( 4 \beta J m \right) \bigr\} - 2 \beta J w' \left( 4 \beta J m \right) \delta m - 4 \left( \beta J \right) ^2 v_{\infty} w'' \left( 4 \beta J m \right) \nonumber \\
& = & - \frac{\delta m }{4} + \frac{\beta J }{2 \cosh ^2 \left( 2 \beta Jm \right) } \cdot \delta m - \frac{ \left( \beta J \right) ^2 \sinh \left( 2 \beta Jm \right) }{\cosh ^3 \left( 2 \beta Jm \right) } \cdot v _{\infty} ,
\label{MF_dynam2body_3a}
\end{eqnarray}
\begin{eqnarray}
& & \frac{1}{4} \frac{d}{dt} \left\{ v _{ij} + 2 m \delta m \right\} \nonumber \\
& = & \frac{\delta m}{2} \left\{ w ( -4 \beta J m ) - w ( 4 \beta J m ) \right\} - \frac{v_{ij} + 2m \delta m }{2} \left\{ w ( 4 \beta J m ) + w ( -4 \beta J m ) \right\}  \nonumber \\
& & - 4 \beta J U \left( j-i \right) w' \left( 4 \beta J m \right) \left( 1- m ^2 \right) - 4 \beta J w' \left( 4 \beta J m \right) ( v _{\infty } + m \delta m ) - 8 \left( \beta J \right) ^2 v_{\infty} w'' \left( 4 \beta J m \right) m \nonumber \\
& = & \frac{\tanh \left( 2 \beta Jm \right) }{2} \delta m - \frac{v_{ij} + 2m \delta m }{2} + \frac{\beta J }{\cosh ^2 \left( 2 \beta Jm \right) } \left\{ U \left( j-i \right) \left( 1- m ^2 \right)  + v _{\infty } + m \delta m \right\} \nonumber \\
& & - \frac{ 2 \left( \beta J \right) ^2 \sinh \left( 2 \beta Jm \right) }{\cosh ^3 \left( 2 \beta Jm \right) } \cdot m v _{\infty}
\label{MF_dynam2body_3b}
\end{eqnarray}

Rearranging (\ref{MF_dynam2body_3a}) and (\ref{MF_dynam2body_3b}), they can be transformed into ordinary differential equations,
\begin{eqnarray}
\frac{ d \left( \delta m \right)}{dt} 
& = & - \delta m + \frac{2\beta J }{ \cosh ^2 \left( 2 \beta Jm \right) } \cdot \delta m - \frac{4 \left( \beta J \right) ^2 \sinh \left( 2 \beta Jm \right) }{\cosh ^3 \left( 2 \beta Jm \right) } \cdot v _{\infty} , \label{MF_dynam2body_3a2} \\
 \frac{dv_{ij} }{dt} & = & - 2 v_{ij} + \frac{4\beta J }{ \cosh ^2 \left( 2 \beta Jm \right) } \cdot \left\{ U \left( j-i \right) \left( 1- m ^2 \right)  + v _{\infty } \right\} . 
\label{MF_dynam2body_3d}
\end{eqnarray}
These equations describe the finite-size effect of time development. As we already explained, some terms appearing in the perturbation of the infinite-range model become the higher-order infinitesimals in the case of our present model. This is why (\ref{MF_dynam2body_3a2}) and (\ref{MF_dynam2body_3d}) have the different forms from the counterparts of the infinite-range model. 
Note that if the system size and $|j-i|$ are sufficiently large, $U(j-i)$ converges to zero. Hence, $v_{\infty}$ is described by the following equation:
\begin{equation}
 \frac{dv_{\infty} }{dt}  = - 2 v_{\infty} + \frac{4\beta J }{ \cosh ^2 \left( 2 \beta Jm \right) } \cdot v _{\infty } . 
\label{MF_dynam2body_3d_2}
\end{equation}

\section{Simulation \label{Simulation}}

\begin{figure}[hbp!]
\begin{center}
\includegraphics[width = 15.0cm]{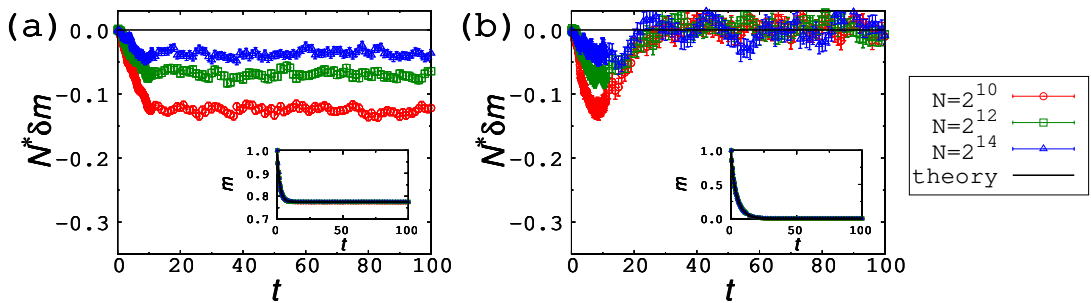} 
\end{center}
\caption{(Color online) Time development of $N^{\ast} \delta m$ for $\alpha = 0.5$ at (a) $T=1.5(<T_c)$ and (b) $T=2.5(>T_c)$. The red circular, green square, and blue triangular points denote the results of MCMC simulations at $N = 2^{10}$, $2^{12}$, and $2^{14}$, respectively, and the black lines are solutions of (\ref{MF_dynam2body_3a2}) and (\ref{MF_dynam2body_3d}). In the insets, the points denote the magnetization calculated by the simulations, and the black curves are the solution of (\ref{MF_dynam_m2}). }
\label{dmsim_05}
\end{figure}

\begin{figure}[hbp!]
\begin{center}
\includegraphics[width = 15.0cm]{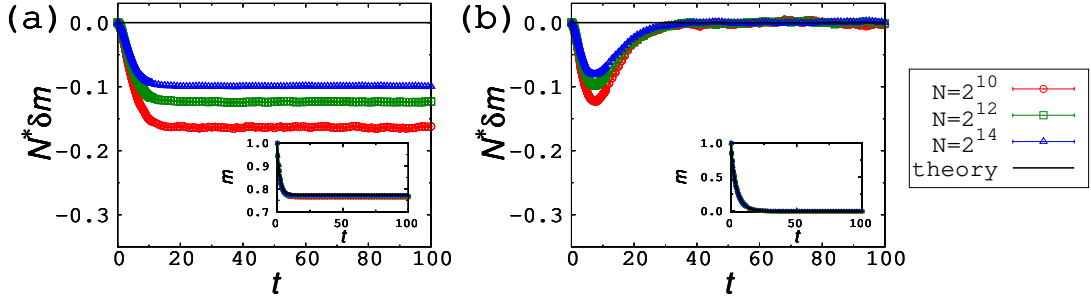} 
\end{center}
\caption{(Color online) Time development of $N^{\ast} \delta m$ for $\alpha = 1$ at (a) $T=1.5(<T_c)$ and (b) $T=2.5(>T_c)$. The meanings of the points and lines are the same as in Figure \ref{dmsim_05}. In the insets, the points denote the magnetization calculated by the simulations, and the black curves are the solution of (\ref{MF_dynam_m2}). }
\label{dmsim_10}
\end{figure}

\begin{figure}[hbp!]
\begin{center}
\includegraphics[width = 15.0cm]{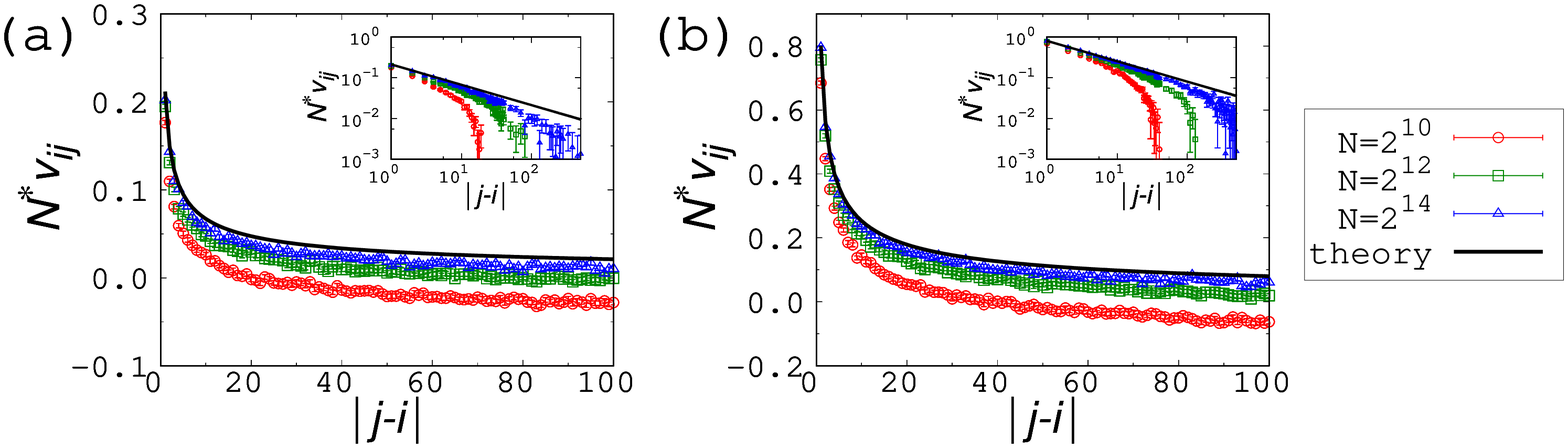} 
\end{center}
\caption{(Color online) Spatial correlation $v_{ij}$ for $\alpha = 0.5$ at $t=100$ and (a) $T=1.5(<T_c)$ and (b) $T=2.5(>T_c)$. The meanings of the points are the same as in Figure \ref{dmsim_05}, and the black curves are solutions of (\ref{MF_dynam2body_3a2}) and (\ref{MF_dynam2body_3d}). The log-log graphs of these data are plotted in the inset. }
\label{scsim_05}
\end{figure}

\begin{figure}[hbp!]
\begin{center}
\includegraphics[width = 15.0cm]{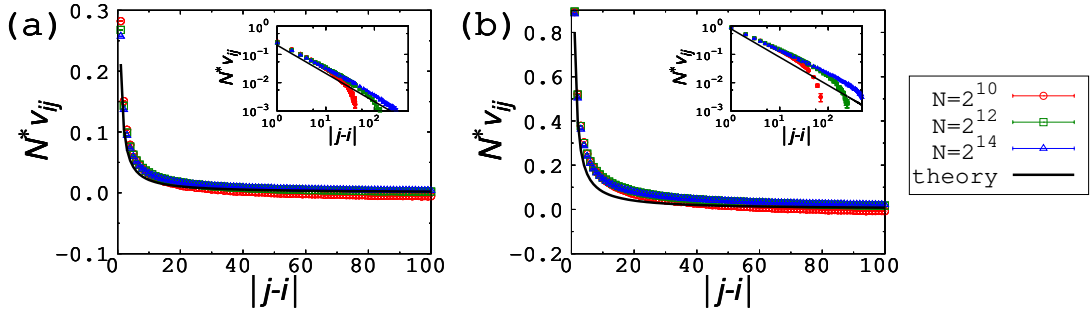} 
\end{center}
\caption{(Color online) Spatial correlation $v_{ij}$ for $\alpha = 1$ at $t=100$ and (a) $T=1.5(<T_c)$ and (b) $T=2.5(>T_c)$. The meanings of the points and curves are the same as in Figure \ref{scsim_05}. The log-log graphs of these data are plotted in the inset. }
\label{scsim_10}
\end{figure}

To investigate the accuracy of the above method, we compared the solutions of (\ref{MF_dynam_m2}), (\ref{MF_dynam2body_3a2}) and (\ref{MF_dynam2body_3d}) with the results of actual MCMC simulations. In these simulations, averages over 192,000 independent trials were taken for each property, and $J$ was fixed as $J = 1$. The initial state was set as the perfectly ferromagnetic state, that is, $m = 1$, and $\delta m = v_{ij} = 0$. Imposing this condition, $\delta m = v_{\infty} = 0$ for arbitrary $t$, if (\ref{MF_dynam2body_3a2}) and (\ref{MF_dynam2body_3d}) are correct. To calculate $v_{ij}$ under these equations, we used the fourth-order Runge--Kutta method with the time interval $\delta t = 1.0 \times 10^{-3}$. Note that in the actual simulations, $\delta m$ was defined as the difference between the calculated magnetization and the theoretical value of $m$ in the thermodynamic limit, obtained using (\ref{MF_dynam_m2}) with the time interval $\delta t = 1/N$.
 The results are shown in Figures \ref{dmsim_05} -- \ref{scsim_10}. Figures \ref{dmsim_05} and \ref{dmsim_10} show time development of $\delta m$ in the case of $\alpha = 0.5$ and $1$, respectively, and the insets of these figures are the comparison between the magnetization itself and $m$ calculated by (\ref{MF_dynam_m2}). We can see from these insets that the time development of the magnetization itself is described by (\ref{MF_dynam_m2}) with high accuracy. Figures \ref{scsim_05} and \ref{scsim_10} show the relation between the spatial correlation $v_{ij}$ and the distance $\left| i - j \right|$ at $t=100$. In each figure, the case in which $T=1.5(<T_c)$ and $T=2.5(>T_c)$ are presented. The insets of Figures \ref{scsim_05} and \ref{scsim_10} show log-log graphs. Note that the horizontal axes of Figures \ref{dmsim_05} -- \ref{dmsim_10} represent the time $t$, while those of Figures \ref{scsim_05} and \ref{scsim_10} are the distance between two spins $\left| i - j \right|$.

\begin{figure}[hbp!]
\begin{center}
\includegraphics[width = 8.0cm]{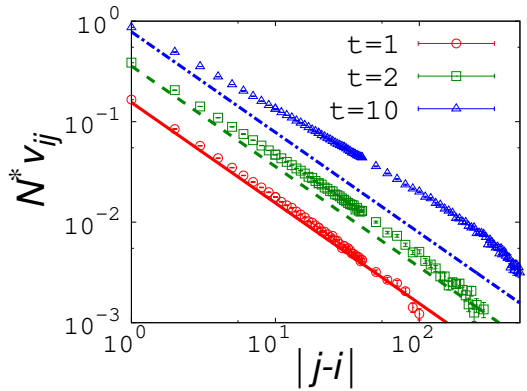} 
\end{center}
\caption{(Color online) Spatial correlation $v_{ij}$ for $\alpha = 1$ and $N=2^{14}$ at $T=2.5$ with different times. The red circular, green square, and blue triangular points denote the results of MCMC simulations at $t = 1$, $2$, and $10$, respectively, and the red solid, green dotted, and blue dash-dotted lines are solutions of (\ref{MF_dynam2body_3a2}) and (\ref{MF_dynam2body_3d}) at the corresponding times. Note that the data at $t=100$, which are plotted in Figure \ref{scsim_10}.(b), are omitted from this graph because they have similar values to those at $t=10$. }
\label{scsim_time}
\end{figure}

As shown in these graphs, $\delta m$ seemed to approach the theoretical evaluation, $\delta m = 0$, with increasing $N$ in every case. The spin correlation $v_{ij}$ also converged to the expectation of (\ref{MF_dynam2body_3a2}) and (\ref{MF_dynam2body_3d}) when $\alpha = 0.5$. However, it exhibited a different behavior from expectation when $\alpha = 1$. Moreover, the time development of $v_{ij}$ are plotted for $\alpha = 1$ and $N=2^{14}$ at $T=2.5$ in Figure \ref{scsim_time}. According to this graph, the difference between the numerical simulation and our approximation was initially small before increasing with increasing $t$. We also calculated similar data at $T=1.5$; however, the graph was omitted from this paper because the tendency was qualitatively similar, that is, the difference increased as time progressed.

For the case in which $\alpha = 1$, convergence to the thermodynamic limit is thought to be slow because the small parameters of perturbation are $O(1/\log N)$. However, as shown in the log-log graphs of Figures \ref{scsim_10} and \ref{scsim_time}, the exponent of the power-law decay itself did not coincide with the theoretical expectation; hence, this disagreement cannot be explained only by slow convergence.

\begin{figure}[hbp!]
\begin{center}
\includegraphics[width = 15.0cm]{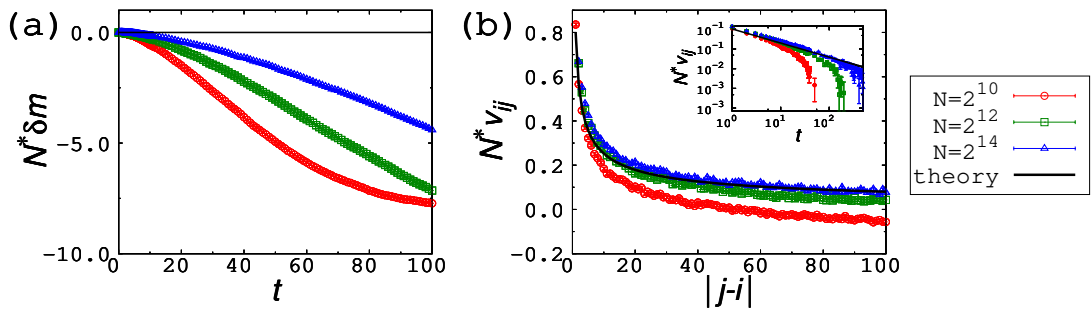} 
\end{center}
\caption{(Color online) Comparison between numerical simulation and our theory at $\alpha = 0.5$ and $T=2.0(=T_c)$. Graphs (a) and (b) represent the time development of $\delta m$ and the spatial correlation $v_{ij}$ at $t=100$, respectively. The meanings of the points, lines, and curves are the same as in Figures \ref{dmsim_05} -- \ref{scsim_10}. The log-log graph of $v_{ij}$ is plotted in the inset of (b). }
\label{Tcsim}
\end{figure}

Note that in the case of $T = T_c(=2)$, magnetization in the equilibrium state did not have a Gaussian distribution. This means that the assumption we used to deal with the BBGKY hierarchy breaks down. If spins are independent of each other in the initial state, our approximation is thought to be valid before the fluctuation of the magnetism grows large. Figure \ref{Tcsim} shows an example of the comparison between the numerical simulation and our approximation at $T=T_c$. For this figure, the initial state and the calculation methods were the same as for Figures \ref{dmsim_05} -- \ref{scsim_10}, and $\alpha$ was taken as $\alpha = 0.5$. From Figure \ref{Tcsim}(b), the difference of $v_{ij}$ from our approximation appears to be small, at least when $t=100$ and the system size $N$ is sufficiently large. However, it is difficult to discuss the $N$-dependence of the $\delta m - t$ graph (Figure \ref{Tcsim}(a)) because the duration over which the approximation is valid at each $N$ is not clear.
As we explained above, $\delta m = 0$ for arbitrary $t$ under the initial condition of this section. An example of the case when $\delta m \neq 0$ is discussed in \ref{App2}.

\section{Summary \label{Summary}}

In this study, we considered the finite-size effect of the one-dimensional long-range Ising model, in which the interaction decays with the power law, $U(j-i) \sim |j-i| ^ {-\alpha}$, expanding our previous study on the infinite-range Ising model. We began with the mean-field approximation and regarded the finite-size effect as perturbation. To deal with the BBGKY hierarchy, we assumed that extensive properties with specific forms had a Gaussian distribution. Within the range of first-order perturbation, this assumption was equivalent to the Kirkwood superposition approximation. After several calculations, we obtained the ordinary differential equations (\ref{MF_dynam2body_3a2}) and (\ref{MF_dynam2body_3d}), which describe the time development of the difference of the magnetization from the mean field approximation, $\delta m$, and the spatial correlation, $v_{ij}$. Comparing the results of these equations with the actual MCMC simulation, the accuracy of our discussion increased when $\alpha$ was small, which suggests that the decay of the interaction is slow. Considering that the zeroth approximation of this study, (\ref{MF_dynam_m2}), is the mean-field approximation ignoring the distance between variables, the point that the spatial correlation $v_{ij}$ can be described by the perturbation is interesting. It is not difficult to generalize the calculations of this study to higher-dimensional space. However, it is not clear whether our discussions can be applied to cases in which the system is inhomogeneous or the time development obeys magnetization-conserving dynamics. This should be investigated in future.

Note that although this study focused on dynamics under the MCMC method, mean-field-like approaches are widely used for long-range interaction systems under other types of dynamics. In the case of a Hamiltonian system, for example, the time development of the probability distribution of one particle is described by the Vlasov equation, which corresponds to (\ref{MF_dynam_1body}) in our study. This is an important equation used not only for the analyses of infinite-range models, such as the Hamiltonian mean-field model \cite{CDFR14,CDR09,YBBDR04}, but also for those of more realistic systems, such as plasmas \cite{SRBG99,BH07} and self-gravitating systems \cite{LPRTB14}. As with (\ref{MF_dynam_1body}), Vlasov equation describes the systems in the thermodynamic limit, and it is known that finite-size systems sometimes show behaviors that are not explained by this equation\cite{BBDRY06}. 
Although the dynamics itself depends on the definition of the model largely, if the mean-field approximation is valid under the thermodynamic limit, the finite-size effect of long-range systems is thought to be regarded as the perturbation from it. Moreover, the Kirkwood superposition approximation expressed as (\ref{alpha3}) and (\ref{alpha4}) does not depend on the concrete forms of the probability distribution. Hence, we can expect that the method of this study can be applied to more general cases such as the Hamiltonian system and help to understand the finite-size effects of them. Whether this generalization is possible should be explored in future studies.

\appendix
\section{Difficulty in the discussion on the validity of the assumption of Section \ref{Calculation} \label{App1} }

In Section \ref{Calculation}, normalized extensive properties with the form $X/N = \sum _i x(\sigma _i)/N $ are assumed to obey the Gaussian distribution. In this appendix, we discuss the difficulty in verifying this assumption, comparing with the case of our previous study which investigated the infinite-range model. 
 In the previous study, we could confirm the validity of this assumption at least in the case of an equilibrium state. Specifically, considering that the expectation value of $X$ at equilibrium state is expressed as
\begin{eqnarray}
\left< X \right> _{\mathrm{eq} } & = & \frac{\mathrm{Tr} _{ \left\{ \sigma _n \right\} } X e^{-\beta H} }{\mathrm{Tr} _{ \left\{ \sigma _n \right\} } e^{-\beta H} } = \lim _{h \rightarrow +0} \frac{\partial}{\partial h} \log \left( \mathrm{Tr} _{ \left\{ \sigma _n \right\} } e^{-\beta H +hX } \right) \nonumber \\
& = & \lim _{h \rightarrow +0} \frac{\partial}{\partial h} \log \left( \frac{ \mathrm{Tr} _{ \left\{ \sigma _n \right\} } e^{-\beta H +hX } }{ \mathrm{Tr} _{ \left\{ \sigma _n \right\} } e^{-\beta H} } \right) =  \lim _{h \rightarrow +0} \frac{\partial}{\partial h} \log \left< e^{hX} \right> _{\mathrm{eq} } ,
 \label{Xeq}
\end{eqnarray}
the following relation holds for any natural number $n$:
\begin{equation}
\frac{1}{N^n} \lim _{h \rightarrow +0} \frac{\partial^{n} }{\partial h^{n} } \log \left< e^{hX} \right> _{\mathrm{eq} } = \frac{1}{N^n} \lim _{h \rightarrow +0} \frac{\partial^{n-1} }{\partial h^{n-1} } \left< X \right> _{\mathrm{eq} } =  O(N^{1-n} ) .
 \label{delXeq}
\end{equation}
Here, $h$ is an external field introduced as the technique of calculation. The left-hand side of (\ref{delXeq}) means the $n$-th order cumulant of $X/N$. Hence this relation shows that the higher-order cumulants are the higher-order infinitesimals, which could be ignored in the previous study. 
However, whether this discussion can be generalized for the long-range model of this study is unclear.
 To consider this difficulty, we deal with the case that $n=3$ as an example. If we do not ignore the infinitesimal that appeared in the right-hand side of (\ref{delXeq}), (\ref{mv2b}) is modified as
\begin{eqnarray}
& & \frac{1}{N^3} \sum _{i,j,k} A _{i,j,k} \nonumber \\
& \equiv & \frac{1}{N^3} \sum _{i,j,k} \sum _{\sigma_i , \sigma_j , \sigma_k} x(\sigma _i) x(\sigma _j) x(\sigma _k) \left\{ \delta p_3 ( \sigma _i , \sigma _j , \sigma _k ; t ) \right. \nonumber \\
& & \left. - 3 \delta p_2 ( \sigma _i , \sigma _j ; t ) p_1 ( \sigma _k , t ) + 3 \delta p_1 ( \sigma _i , t ) p_1 ( \sigma _j , t ) p_1 ( \sigma _k , t ) + O \left( \frac{1}{N^{\ast 2}} \right) \right\} = O \left( \frac{1}{N^2} \right) . \nonumber \\
 \label{mv2b_N}
\end{eqnarray}
To rationalize (\ref{alpha3}) and following calculations of the main text, $A_{i,j,k}$ itself should be ignorable.
 In the case of the infinite-range model, $A_{i,j,k}$ is the constant independent of  $i,j$, and $k$. Hence (\ref{mv2b_N}) shows that this quantity itself is the higher-order infinitesimal: $A_{i,j,k} = O(1/N^2)$. In the case of the long-range model of this study, on the other hand, it is difficult to prove that $A_{i,j,k}$ itself is the higher-order infinitesimal which can be ignored. 
Hence, to verify the validity of the assumption, the comparison between the numerical simulation discussed in Section \ref{Simulation} is required.

\section{The case that $\delta m \neq 0$ \label{App2} }

Calculations of Section \ref{Simulation} dealt with the case that $\delta m = 0$ for arbitrary $t$. In this appendix, we investigate an example that $\delta m$ has nonzero value. Specifically, we consider the initial state in which the values of spins are given as $\pm 1$ with probability $\left( 1 \pm N^{\ast -1}\right)/2 $, independent of each other. It means that the initial condition is expressed as $m = v_{ij} =0$ and $\delta m = 1/N^{\ast}$. Note that we let the initial magnetization depend on $N^{\ast}$, because $\delta m = O(1/N^{\ast})$. Under this initial condition, $m = v_{\infty} = 0$ for arbitrary $t$ because of (\ref{MF_dynam_m2}) and (\ref{MF_dynam2body_3d_2}). Hence, (\ref{MF_dynam2body_3a2}) is expressed as
\begin{equation}
\frac{ d \left( \delta m \right)}{dt} = - (1- 2 \beta J)\delta m . \label{MF_dynam2body_3a_m0} 
\end{equation}
The solution of (\ref{MF_dynam2body_3a_m0}) under the above initial condition is given as
 \begin{equation}
\delta m = \frac{e^{ - (1- 2 \beta J)t} }{N^{\ast} } . \label{delm_App2} 
\end{equation}

\begin{figure}[hbp!]
\begin{center}
\includegraphics[width = 15.0cm]{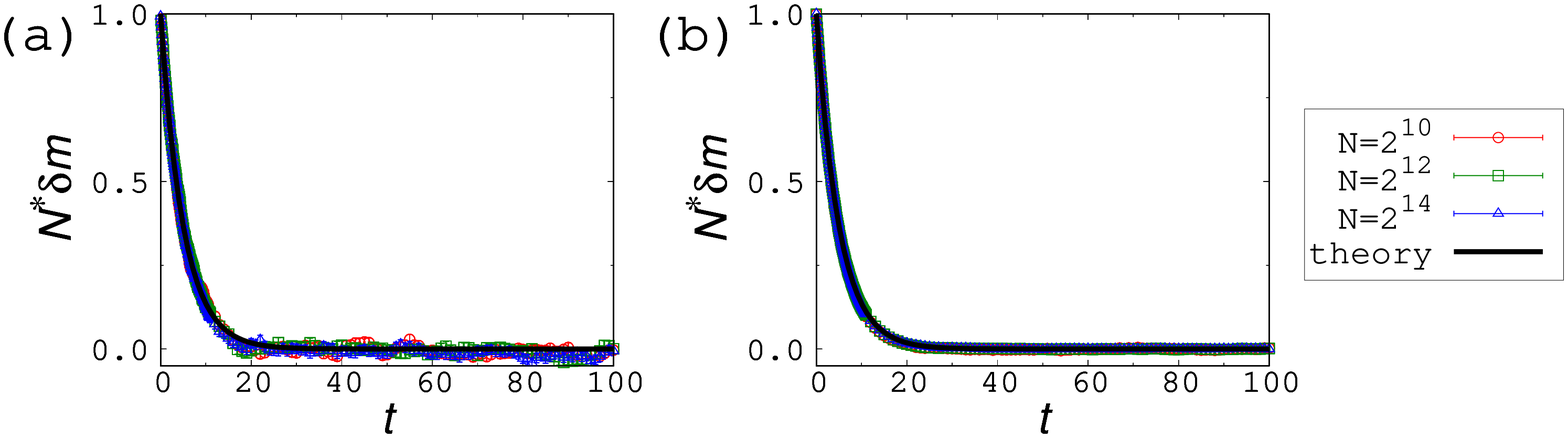} 
\end{center}
\caption{(Color online) Time development of $N^{\ast} \delta m$ for (a) $\alpha = 0.5$ and (b) $\alpha = 1$ at $T=2.5(>T_c)$. The initial condition is given as $m = v_{ij} =0$ and $\delta m = 1/N^{\ast}$. The meanings of the points are the same as in Figures \ref{dmsim_05} -- \ref{scsim_10}, and the black curves indicate (\ref{delm_App2}). }
\label{dmsim_deltastart_T25}
\end{figure}

Figure \ref{dmsim_deltastart_T25} shows the comparison between (\ref{delm_App2}) and the results of the simulations at $T=2.5(>T_c)$. In the simulation, calculation methods except the initial condition are the same as those of Section \ref{Simulation}. Particularly, $\delta m$ is defined as the difference between the calculated magnetization and the solution of (\ref{MF_dynam_m2}), $m$, as in Section \ref{Simulation}. Hence, under the initial condition of this appendix, $\delta m$ coincides with the calculated magnetization itself because $m=0$. Seeing Figure \ref{dmsim_deltastart_T25}, (\ref{delm_App2}) coincides with the behavior of actual system with high accuracy. It means that (\ref{MF_dynam2body_3a2}) actually describes the time development of $\delta m$.

\begin{figure}[hbp!]
\begin{center}
\includegraphics[width = 9.0cm]{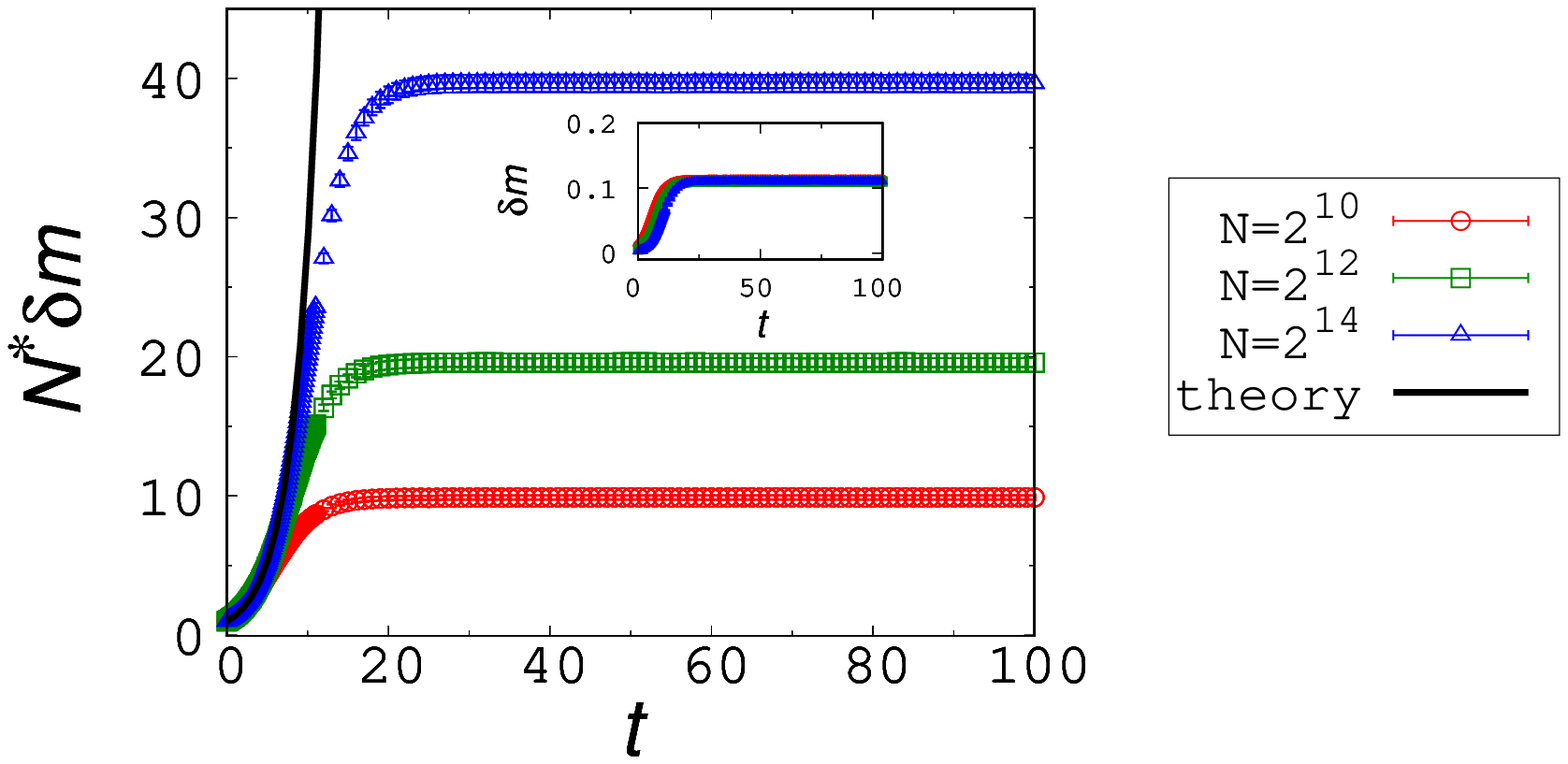} 
\end{center}
\caption{(Color online) Time development of $N^{\ast} \delta m$ for $\alpha = 0.5$ at $T=1.5(<T_c)$. The initial condition and the meanings of the points and curves are the same as in Figure \ref{dmsim_deltastart_T25}. The inset shows $\delta m$ itself, which coincides with the magnetization under the condition of this appendix.  }
\label{dmsim_deltastart_T15}
\end{figure}

Note that in the case that $T < T_c$, $\delta m$ given by (\ref{delm_App2}) diverges as the time passes. This divergence is related to the point that $m=0$ is the unstable solution of (\ref{MF_dynam_m2}). Furthermore, the weak magnetization at $T < T_c$ results in the magnetization reversal, which causes the breakdown of the approximation of this study. Indeed, as we can see from Figure \ref{dmsim_deltastart_T15}, (\ref{delm_App2}) cannot describe the behavior of the simulation except when $t$ is small. In the large-$t$ limit, the magnetization converges to the equilibrium value. Hence, $\delta m$, which is equal to the simulated magnetization under the condition of this appendix, also converges to the asymptotic value. Note that this asymptotic value can be both positive and negative if the initial magnetization is small. Hence, its average over many trials does not simply coincide with the stable fixed point of (\ref{MF_dynam_m2}). This is why the insets of Figures \ref{dmsim_05}.(a) and \ref{dmsim_deltastart_T15}, both of which indicate the magnetization under the same $\alpha$ and $T$, have the different asymptotic values. The breakdown of the approximation caused by the magnetization reversal itself is a serious problem that is also observed in the infinite-range model of our previous study. However, we do not discuss it further because the aim of this appendix is the verification of (\ref{MF_dynam2body_3a2}) under the condition that such problems do not exist. 

\section*{Acknowledgments}
This study was supported by JSPS KAKENHI, Grant Number JP21K13857. 
We would like to thank Editage (www.editage.com) for English language editing.

\end{document}